\documentclass[
reprint,
aps,
amsmath,
amssymb,
]{revtex4-1}

\usepackage{graphicx}
\usepackage{dcolumn}
\usepackage{bm}
\usepackage{color}

\graphicspath{{./fig1/}{./fig2/}{./fig3/}{./fig4/}}

\begin{document}

\title{Optomechanical microwave amplification without mechanical amplification}

\author{Martijn A. Cohen*}
\author{Daniel Bothner*}
\author{Yaroslav M. Blanter}
\author{Gary A. Steele}

\affiliation{Kavli Institute of Nanoscience, Delft University of Technology, 2628 CJ Delft, The Netherlands}
\date{\today}

\begin{abstract}
High-gain and low-noise signal amplification is a valuable tool in various cryogenic microwave experiments. A microwave optomechanical device, in which a vibrating capacitor modulates the frequency of a microwave cavity, is one technique that is able to amplify microwave signals with high gain and large dynamical range. Such optomechanical amplifiers typically rely on strong backaction of microwave photons on the mechanical mode achieved in the sideband-resolved limit of optomechanics. Here, we observe microwave amplification in an optomechanical cavity in the extremely unresolved sideband limit. A large gain is observed for any detuning of the single pump tone within the cavity linewidth, a clear indication that the amplification is not induced by dynamical backaction. By being able to amplify for any detuning of the pump signal, the amplification center frequency can be tuned over the entire range of the broad cavity linewidth. Additionally, by providing microwave amplification without mechanical amplification, we predict that using this scheme it is possible to achieve near-quantum-limited microwave amplification despite a large thermal occupation of the mechanical mode. 
\end{abstract}
\maketitle

\begin{figure}
	\includegraphics{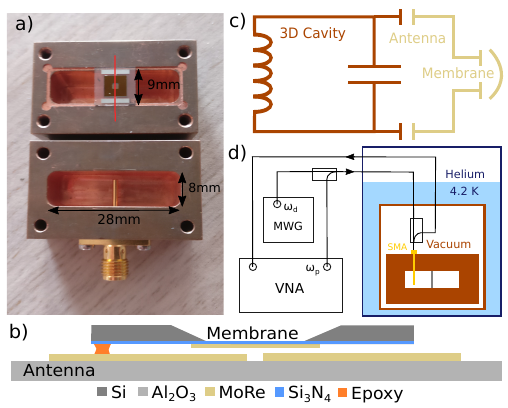}
	\caption{\label{fig1} A 3D optomechanical cavity in the sideband-unresolved regime at 4K. (a) A photograph of the copper 3D cavity used in this experiment. The inner dimensions of the cavity correspond to $28\times28\times8$~mm$^3$ and microwave signals are coupled in an out using a single SMA connector. The depth of the pin determines the external coupling. The red line through the chip shows the cross-section shown in (b). This cross-section shows the thin film layers of the antenna chip and membrane chip. The membrane chip is a 50~nm thick silicon nitride window with 20~nm thick MoRe alloy sputtered on top. The antenna chip is made from double side-polished sapphire with 100~nm thick MoRe antennas patterned on top. The two chips are fixed together with a single drop of epoxy in one corner. (c) A circuit diagram which corresponds to the optomechanical system. The fundamental mode of the 3D cavity is represented by an LC resonator, and the antennas act as capacitors which concentrate the electric field towards the membrane, which itself can be represented as a mechanically compliable capacitor. The membrane motion modulates the resonance frequency of the 3D cavity. (d) We put the 3D optomechanical setup inside a vacuum can and cool it to 4.2~K in liquid helium. Using two directional couplers we probe the cavity by means of a vector network analyzer (VNA) while simultaneously sending a drive signal with a microwave generator (MWG).}
\end{figure}

\section{Introduction}

Amplification is an essential part of any measurement system where there is a need to distinguish a signal from noise. Cryogenic measurement systems for quantum experiments often use a high-electron-mobility transistor amplifier in their amplification chain~\cite{wallraff2004strong, regal2008measuring, teufel2011sideband}. However, they typically operate with higher added noise levels than the theoretical limit imposed by quantum mechanics, something readily achieved in the optical domain~\cite{clerk2010introduction}. Josephson junction based microwave amplifiers~\cite{roy2016introduction, movshovich1990observation}, on the other hand, can have quantum-limited noise and have been used to entangle superconducting qubits~\cite{riste2013deterministic}, to convert quantum states to mechanical motion~\cite{reed2017faithful}, and to implement error-correction in quantum circuits~\cite{ofek2016extending}. In such amplifiers, a Josephson junction is used as a low-loss nonlinear element that enables parametric amplification driven by an external pump tone. 

A mechanically compliable capacitor coupled through radiation pressure to a superconducting circuit~\cite{aspelmeyer2014cavity, regal2008measuring} -- a microwave optomechanical system -- can also be used as a nonlinear circuit and can create a microwave amplifier~\cite{massel2011microwave}. A Josephson parametric amplifier (JPA) typically has a strong $x^3$ Kerr (Duffing) nonlinearity in the restoring force in the equation of motion for parametric amplification (where the coordinate $x$ for a JPA would refer to the phase difference across the junction). Although the Kerr nonlinearity is not necessary for amplification, it can result in amplifier saturation already at 100 photons~\cite{eichler2014controlling, ockeloen2016low}.

In optomechanical amplification cubic nonlinearities are weak compared to a JPA. Typical schemes operate by driving the optomechanical system at a frequency positively detuned from the cavity frequency (blue-sideband driving): doing so, one can produce amplification analogous to a non-degenerate two-mode amplifier, where the cavity mode and mechanical mode act as signal and idler, respectively~\cite{massel2011microwave}. With mechanical frequencies in the MHz range, it is then neccessary to cool the mechanical element to $<$~50~$\mu$K such that the microwave amplification is not dominated by thermal noise and the amplifier can approach the quantum limit. A more recent optomechanical microwave amplification scheme has approached this problem by using two pump tones, a red-detuned tone to provide cooling to the mechanical mode and a blue-detuned one to provide amplification, either in one cavity~\cite{ockeloen2017noiseless, ockeloen2017theory}, or in two separate cavity modes coupled to one mechanical mode~\cite{ockeloen2016low}. 

Here, we present an observation of amplification of cavity fields mediated by a mechanical oscillator which, in contrast to earlier works~\cite{massel2011microwave}, does not make use of dynamical backaction, and results in no amplification of the mechanical motion.
Strikingly, we observe optomechanical microwave amplification in the presence of a drive tone red-detuned from the cavity resonance frequency, a regime associated with mechanical damping and not amplification.
The amplification mechanism observed does not rely on dynamical backaction, and results in no amplification of the mechanical motion.
Based on the mechanism we identify, we predict that the amplification presented here has the potential to amplify microwave signals with near-quantum-limited added noise, even in the presence of large thermal occupations of the mechanical resonator without cooling of the mechanical mode.
We note that the basis for this single-tone amplification method has been identified in previous theoretical works \cite{botter2012linear, kamal2017minimal} but has not been explored experimentally.

\begin{figure}
	\includegraphics{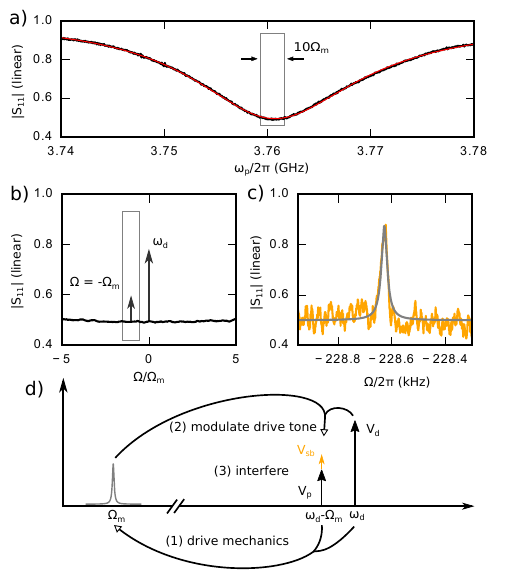}
	\caption{\label{fig2} Optomechanically induced reflection (OMIR) in the sideband-unresolved regime. (a) A reflection measurement of the copper cavity, black dots show raw data and red line a fit. The fit result in a center frequency of $\omega_0 / 2\pi = 3.76$~GHz and a linewidth of $\kappa / 2\pi = 23.5$~MHz. The grey rectangle is zoomed in on panel (b) and shows the width of 10 mechanical frequencies, $10\Omega_\textrm{m}$, a visual demonstration that the optomechanical system is deep in the sideband-unresolved regime. The curve is normalized to a background level detuned from the resonance. (b) Experimental scheme: We send a strong drive tone at $\omega_\textrm{d} = \omega_0~(\Delta = 0) $ and sweep a weak additional probe tone around one mechanical frequency detuned from the drive tone, here $ \omega_\textrm{p} - \omega_\textrm{d} = \Omega \approx -\Omega_\textrm{m}$. The grey box is zoomed-in in the next subfigure. (c) A sample reflection measurement that we obtain when driving at low power. What results is an OMIR interference effect. (d) A diagram which shows the basic mechanism of the observed effect. (1) The probe and the drive tone interfere, causing a beating pattern which oscillates at the mechanical frequency. By means of radiation pressure, this coherently drives the mechanical resonator. (2) The coherent oscillations of the mechanical oscillator then in turn modulate the cavity frequency which phase modulates the drive tone, creating sidebands at $\pm \Omega_\textrm{m}$. (3) The sideband and the probe tone interfere, which gives rise to OMIR. Note that there is no cooling or amplification of the mechanical oscillator when $\Delta = 0$.}
\end{figure}

\section{Experimental Setup}

The experimental setup is shown in Fig.~\ref{fig1}. As microwave cavity we use a 3D copper cavity with dimensions $28 \times 28 \times 8$~mm$^3$ with a bare frequency of $7.59$~GHz. It contains a stacked microchip structure with a mechanical oscillator, capacitively coupled to the cavity. The mechanical capacitor is constructed using a flip-chip technique where the antennas and membrane are fabricated separately and joined together using an adhesive~\cite{yuan2015large, noguchi2016ground}. The antenna chip contains two conducting strips which concentrate the electric field, and is fabricated on a double-side polished sapphire chip with superconducting Molybdenum-Rhenium (MoRe) 60-40 alloy electrodes~\cite{singh2014molybdenum}. The membrane chip is a square $5 \times 5$~mm transmission-electron microscopy window from Norcada made from 50~nm thick stoichiometric Si$_3$N$_4$~\cite{zwickl2008high} on which we sputter a 20~nm thick MoRe square patch. A discussion on the chosen metallization geometry is given in the Supplemental Material \cite{yu2012control, cohen2019SM}. The two chips are then fixed together using a single drop of epoxy in the corner of the membrane chip. The gap between two chips is measured at room temperature to be  2~$\mu$m using the depth-of-focus of an optical microscope.

The microwave optomechanical system can be understood using the effective circuit diagram in Fig.~\ref{fig1}(c). The fundamental mode of the 3D cavity is represented by an LC circuit, and the antenna-chip structure is a capacitive circuit in which the middle two capacitors are modulated in-phase by the motion of the membrane. We measure the 3D optomechanical setup in vacuum with a bath temperature of 4.2~K by means of a microwave reflection measurement using an Agilent PNA N5221A vector network analyzer (VNA). In the case that  measurements require two microwave tones, a weak probing tone, with frequency $\omega_\textrm{p}$, is provided by the output port of our VNA, and a second stronger drive tone, with frequency $\omega_\textrm{d}$, is generated by a Rohde \& Schwarz SMB100A microwave generator. The signals are combined using a directional coupler where the transmitted port is used for the drive tone to allow for maximum drive power.

\begin{figure}
	\includegraphics{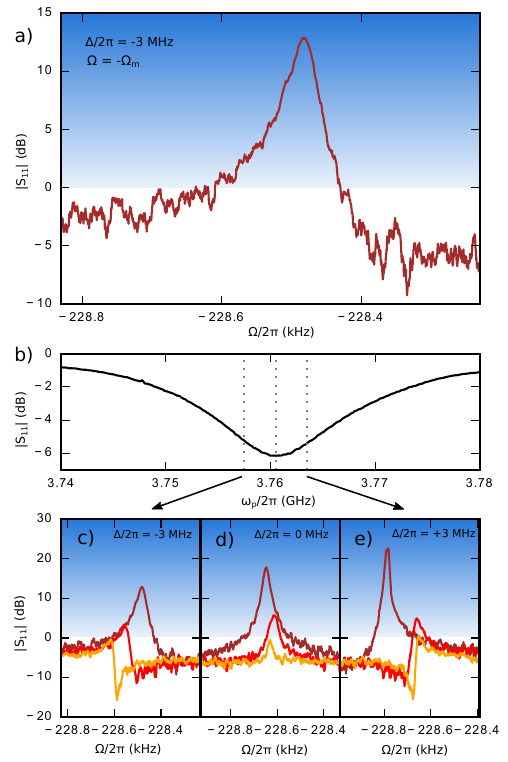}
	\caption{\label{fig3} Optomechanical microwave amplification with negative probe detuning. (a) A reflection measurement when the drive detuning $\Delta / 2\pi = -3$~MHz at -11~dBm, and probe detuning $\Omega = \omega_\textrm{p} - \omega_\textrm{d} = -\Omega_\textrm{m}$ showing gain up to 13~dB. (b) The experimentally determined cavity resonance with three vertical lines indicate the positions for the three drive detunings $\Delta_1 / 2\pi = -3$~MHz, $\Delta_2 / 2\pi = 0$~MHz, $\Delta_3 / 2\pi = +3$~MHz used in (c), (d), (e), respectively. (c)-(e) Probe sweeps of the lower sideband $\omega_\textrm{p} \approx \omega_0 - \Delta_\textrm{i} - \Omega_\textrm{m}$ for three different drive tone powers, -11~dBm (brown), -15~dBm (red), -19~dBm (orange), showing probe tone gain for all three chosen detunings. The cavity resonance shift slightly with respect to drive power and detuning. The responses are normalized such that a signal above 0~dB corresponds to gain $>$ 1. The highest gain is observed when the drive detuning is positive, which shows that dynamical backaction does play a contributing role. One can also see an asymmetric Fano lineshape when the detuning is non-zero. }
\end{figure}

\section{Results}

\subsection{OMIR in the Sideband-Unresolved Regime}

Figure \ref{fig2} shows a characterization of the optomechanical system. The cavity resonant frequency is found to be $\omega_0 / 2\pi = 3.76$~GHz, with a linewidth of $\kappa / 2\pi = 23.5$~MHz, (intrinsic losses $\kappa_\textrm{i}/2\pi = 17.6$~MHz, external losses $\kappa_\textrm{e}/2\pi = 5.9$~MHz) which results in a relatively low quality factor of $Q = 160$ compared to other microwave optomechanical amplification devices, which we attribute to quasiparticle losses of the thinner MoRe film on the membrane. This hypothesis is supported by the observation of higher internal quality factors when measuring the cavity with only the antenna chip. The low quality factor benefits us, however, as this increases the tuning range of our amplification. 

A reflection measurement of the cavity resonance including a fit line is shown in Fig.~\ref{fig2}(a). The cavity is undercoupled, as $\eta = \kappa_\textrm{e} / \kappa = 0.25$.
To measure the mechanical resonance, a two-tone measurement scheme called optomechanically induced reflection (OMIR), an analogue of optomechanically induced transparency~\cite{weis2010optomechanically} in a reflection geometry is used, illustrated in Fig.~\ref{fig2}(b). 
A strong drive tone is applied at the cavity resonance frequency, and a second weak probe tone is swept around a range of frequencies detuned by the mechanical frequency from the drive tone.
We define a detuning $\Delta = \omega_\textrm{d} - \omega_0$ between the drive frequency $\omega_\textrm{d}$ and the cavity resonance frequency, as well as a detuning  $\Omega = \omega_\textrm{p} - \omega_\textrm{d}$ between the drive frequency and the probe frequency $\omega_\textrm{p}$. When $\Omega = \pm\Omega_\textrm{m}$, an interference effect in the measured reflection at the probe frequency is observed, shown in Fig.~\ref{fig2} for the case where $\Omega = -\Omega_\textrm{m}$. In the measurement, the mechanical resonance is excited by the oscillating radiation pressure from the beating of the pump and drive tones, which then creates a sideband of the drive tone which interferes with the probe field. Using this technique, we find a mechanical frequency of $\Omega_\textrm{m}/2\pi = 228.65$~kHz and damping rate $\gamma_m / 2\pi = 22.0$~Hz. From the ratio $\kappa / \Omega_\textrm{m} = 103$, the system is found to be deep in the sideband-unresolved limit, illustrated in Fig.~\ref{fig2}(a) and (b). We also estimate our coupling constant $g_0 = 69$~mHz using finite element simulations (see Supplemental Material Sec.~2 \cite{cohen2019SM}).

\begin{figure}
	\includegraphics{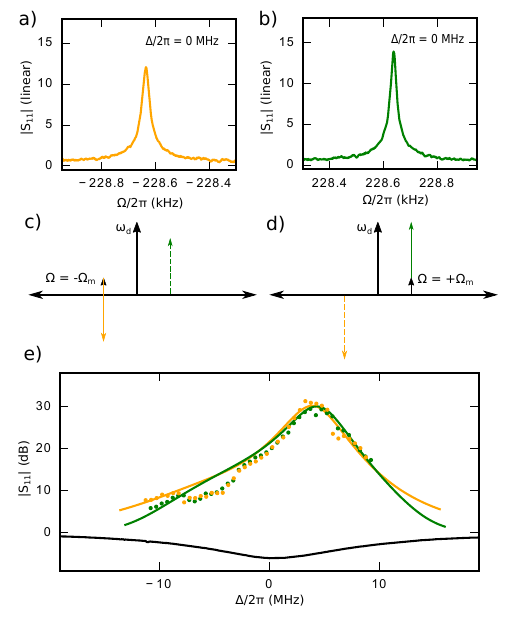}
	\caption{\label{fig4} Understanding the mechanism of amplification and how the gain depends on probe and drive detuning. (a)-(b) Reflection spectra taken with a -9~dBm drive on cavity resonance and with the probe at the lower drive sideband $\Omega = -\Omega_\textrm{m}$ in (a), and the probe at the upper drive sideband $\Omega = +\Omega_\textrm{m}$ in (b). Note that the maximum gain differs between the sidebands. (c)-(d) A more precise explanation of the interference effect, idealized for when the drive detuning is zero, $\Delta = 0$. The phase modulations of the drive tone produce sidebands which have opposite phase and thus interfere constructively or destructively with the probe tone depending on which sideband is probed. This is why one gets a slightly lower gain for $\Omega = -\Omega_\textrm{m}$ (c) and slightly higher gain for $\Omega = +\Omega_\textrm{m}$ (d). (e) We performed a sweep of varying drive detunings, $\Delta$, and measured the maximum gain value at constant drive power for both the upper and lower drive sidebands, in yellow and green, respectively. Dots show experimental data points and lines show the result of theoretical calculations. As one might expect, the gain is the highest when there is a positive contribution from dynamical backaction. We note here, that in contrast to experiments in the sideband-resolved limit, we can observe OMIR on both sides of the drive tone for essentially all detunings. In particular for the case of a resonant drive $\Delta = 0$, the simultaneous generation of both sidebands leads to a cancellation of the phase-lagging and phase-leading sideband backaction forces, independent of the magnitude of these sidebands. In combination, these effects allow for microwave amplification without amplification of the mechanical motion.}
\end{figure}

\subsection{Optomechanical microwave amplification without dynamical backaction}

Figure \ref{fig3}(a) shows a measurement similar to that in Fig.~\ref{fig2}(c) but now with $\Delta/2\pi = -3$~MHz and higher drive power. Strikingly, we observe that the reflection coefficient goes significantly above one, indicating that there is microwave amplification of 13~dB being performed by the system. This is surprising since in the usual paradigm in optomechanics, a negative detuning results in damping and not amplification: a strong indication that dynamical backaction is not the origin of the observed microwave gain.

We can understand the origin of the observed microwave amplification qualitatively using the illustration in Fig.~\ref{fig2}(d) and looking at how the amplitudes of the different signals change when increasing the amplitude of the drive tone $V_\textrm{d}$ at $\omega_\textrm{d}$ while keeping the amplitude of the probe signal $V_\textrm{p}$ constant. The amplitude of the mechanical motion, indicated by the height of the arrow $\Omega_\textrm{m}$ is proportional to the product of the probe and the drive tone amplitudes; increasing the drive power will drive the mechanical resonator to a large coherent amplitude. The drive tone mechanical sideband amplitude $V_\textrm{sb}$ has a height that is proportional to both the mechanical amplitude and the drive amplitude $V_\textrm{d}$. Consequently, the sideband of the drive, $V_\textrm{sb}$, will be proportional to $V_\textrm{d}^2$. For sufficiently large drive powers, $V_\textrm{sb}$ will become larger than $V_\textrm{p}$: if they add in phase, this will then give maximum amplification.

One way to think about this amplification process is as a frequency mixing process, where the drive tone is not only down-converting the signal, but also amplifies it. In this sense, this amplification process can be thought of as a ``double mixer amplifier'' (see Supplemental Material Sec.~6 \cite{cohen2019SM}). Note that this ``double mixer'' process does not make use of dynamical backaction~\cite{aspelmeyer2014cavity}, and also does not result in any mechanical amplification. For $\Delta = 0$, there is only coherent driving of the mechanical oscillator with no damping or amplification, while when $\Delta < 0$ both mechanical cooling and microwave amplification are attained simultaneously with a single drive tone. The fact that microwave amplification can occur without mechanical amplification also potentially enables near-quantum-limited amplification even if the mechanical resonator is not cooled to the ground state. The near-quantum-limited regime is reached in the optomechanical mixing amplification scheme when  $V_\textrm{d}$ is large enough such that the first step of the mixing amplification, corresponding to the radiation pressure driving of the mechanical resonator, results in translation of the quantum noise of the input probe field to an amplitude that is larger than the thermal noise of the mechanical mode. Quantum limited operation becomes possible for cooperativities $C >  kT_\textrm{m} / \hbar \Omega_\textrm{m} = n_\textrm{th}$ for an optimal amplifier configuration, corresponding to the criteria of reaching the radiation pressure shot noise limit where the quantum fluctuation of the input field dominate the force noise of the mechanical resonator \cite{teufel2016overwhelming}. Furthermore, our amplification mechanism can be seen as a specific case of non-degenerate parametric amplification, where the strong drive and probe tones act as the pump and signal tones, respectively. The way our process differs from previously studied optomechanical microwave amplification techniques is that the idler tone is the other sideband of the drive tone, in contrast to the single-tone blue-sideband amplification case, which can be also be understood as a non-degenerate parametric amplification scheme but in which the mechanical mode plays the role of the idler (see Supplemental Material Sec.~7 \cite{cohen2019SM} for further discussion).

Figure \ref{fig3}(c)-(e) show the microwave responses for different detunings $\Delta$, illustrated in Fig.~\ref{fig3}(b). We observe that the gain depends on detuning, with the largest amplification occurring for positive $\Delta$. This indicates that there is also a contribution from dynamical backaction in these measurements, reducing the gain for $\Delta < 0$ and providing additional gain for $\Delta > 0$. The data also show a shift of the mechanical frequency for $\Delta \neq 0$ due to the optical spring effect, confirming the presence of dynamical backaction. We note however that this dynamical backaction is not needed for the amplification: amplification also occurs at $\Delta = 0$ in the absence of dynamical backaction, and for $\Delta < 0$ in spite of mechanical damping from dynamical backaction.

In Fig.~\ref{fig4}, we explore in detail the dependence of the gain on detuning $\Delta$ of the drive tone from the cavity resonance, for both positive and negative $\Omega = \pm \Omega_\textrm{m}$, respectively. Note that in the sideband-resolved limit $\kappa < \Omega_\textrm{m}$, features corresponding to $\Omega$ and $\Delta$ having the same sign are usually not accessible experimentally as they correspond to features far outside the cavity resonance. However in the case $\kappa \gg \Omega_\textrm{m}$, both of these features are accessible and can be equally strong. Figure~\ref{fig4}(a) and (b) show the reflection coefficient of the cavity for $\Delta=0$ and for positive and negative $\Omega$, respectively. It is interesting to note that the gain is larger for $+\Omega_\textrm{m}$ than for  $-\Omega_\textrm{m}$. This can be understood by the fact that with a drive on cavity resonance the up-converted mechanical sidebands of the drive tone have opposite phase compared to the probe signal, as shown in (c) and (d). This asymmetry in (a) and (b) can then be understood as arising from the difference in constructive and destructive interference of the probe and the sideband, as shown in (c) and (d). Fig.~\ref{fig4}(e) shows the dependence of the observed and theoretically calculated gain for positive (green) and negative (orange) $\Omega$ as a function of drive-cavity detuning $\Delta$. Due to the additional gain from dynamical backaction, the gain is maximum for positive drive detuning, but is still larger than unity for all detunings when the drive is sufficiently strong. We also note that there are several transitions where the relative amplitudes of positive and negative $\Omega = \pm \Omega_\textrm{m}$ changes sign (crossing of the green/yellow curves in Fig.~\ref{fig4}(e)), which is a result of changes in the relative phase of the sidebands as a function of detuning (see Supplemental Material Sec.~6 \cite{cohen2019SM} for further discussion, as well as a notebook to calculate the reflection coefficient curves using the optomechanical equations of motion \cite{cohen2019code}).
\section{Conclusions}

In conclusion, we have demonstrated microwave amplification in an optomechanical system which does not depend on dynamical backaction. Since our amplification scheme works even in the sideband-unresolved limit, we are able to center the amplification window over a relatively wide cavity linewidth. Furthermore, since the thermal noise of the mechanical mode is not amplified, this method could achieve near-quantum-limited microwave amplification without the necessity to cool the mechanical oscillator.

\subsection*{Acknowledgements}

This project has received funding from the European Union’s Horizon 2020 research and innovation program under Grant Agreement No. 732894 - Hybrid Optomechanical Technologies, from the European Research Council (ERC) under the European Union’s Horizon 2020 research and innovation program (Grant Agreement No. 681476 - Quantum Optomechanics in 3D), and was supported by the Netherlands Organisation for Scientific Research (NWO) in the Innovational Research Incentives Scheme – Vidi, project 680-47-526.

\end{document}